\documentclass[journal]{IEEEtran}
\usepackage[cmex10]{amsmath}
\usepackage{siunitx}
\usepackage[font=footnotesize]{caption}
\usepackage{psfrag}
\usepackage[utf8]{inputenc}
\usepackage[T1]{fontenc}
\usepackage{amsmath,amsfonts,amsbsy,amssymb}
\usepackage{mathabx}
\usepackage{mathrsfs}
\usepackage[nolist]{acronym}
\usepackage{tabularx}
\usepackage{amssymb}
\usepackage{relsize}
\usepackage{pdfpages}
\usepackage{amsmath}
\usepackage[export]{adjustbox}
\usepackage{graphicx}
\usepackage{soul}
\usepackage{cite}
\usepackage{multirow}
\usepackage{wasysym}
\usepackage{multirow}
\usepackage{float}
\usepackage{xcolor}
\usepackage{subfig}
\usepackage[margin=16mm]{geometry}
\usepackage{ragged2e}
\usepackage{makecell, multirow, tabularx}

\usepackage{siunitx, mhchem}
\usepackage{xcolor}
 
\usepackage[ruled, lined, longend, linesnumbered]{algorithm2e}

\usepackage{xcolor}
\usepackage{url}
\usepackage{multicol}
\usepackage{verbatim}
\usepackage{array}
\definecolor{uranianblue}{rgb}{0.69,0.86,0.96}
\definecolor{thistle}{rgb}{0.85,0.75,0.85}
\definecolor{mintgreen}{rgb}{0.6,1.0,0.6}
\definecolor{new}{rgb}{0.4940 0.1840 0.5560}
\definecolor{azurexwebcolor}{rgb}{0.94,1.0,1.0}

\title{Home Energy Management Systems: Operation and Resilience of Heuristics against Cyberattacks}
\author{Hafiz Majid Hussain, Arun Narayanan, Subham Sahoo, Yongheng Yang, Pedro H. J. Nardelli and Frede Blaabjerg\thanks{PHJN, HMH and AN are with LUT University, Finland. SS and FB are with Department of Energy, Aalborg  University,  Denmark. YY is with Department of Electrical Engineering, Zhejiang University. China. This paper is partly supported by Academy of Finland via: (a) ee-IoT project n.319009, (b) FIREMAN consortium CHIST-ERA/n.326270 (CHIST-ERA-17-BDSI-003), and (c) EnergyNet Research Fellowship n.321265/n.328869.}}

\begin{document}

\maketitle
\thispagestyle{empty}
\pagestyle{empty}

\begin{abstract}

Internet of Things (IoT) and advanced communication technologies have demonstrated great potential to manage residential energy resources by enabling demand-side management (DSM). Home energy management systems (HEMSs) can automatically control electricity production and usage inside homes using DSM techniques. These HEMSs will wirelessly collect information from hardware installed in the power system and in homes with the objective to intelligently and efficiently optimize electricity usage and minimize costs. However, HEMSs can be vulnerable to cyberattacks that target the electricity pricing model. The cyberattacker manipulates the pricing information collected by a customer's HEMS to misguide its algorithms toward non-optimal solutions. The customer's electricity bill increases, and additional peaks are created without being detected by the system operator. This article introduces demand-response (DR)-based DSM in HEMSs and discusses DR optimization using heuristic algorithms. Moreover, it discusses the possibilities and impacts of cyberattacks, their effectiveness, and the degree of resilience of heuristic algorithms against cyberattacks. This article also opens research questions and shows prospective directions.

\end{abstract}
%



\section{Introduction}\label{HEMS}

Smart grid technologies and smart meters have enabled customers to know their demand profiles in greater detail while helping electricity grid operators to improve the efficiency and reliability of the power system \cite{01}.
This encourages both customers and grid operators to modify load energy demand profiles to achieve different objectives such as optimizing the usage of renewable energy, reducing peak loads, or moving some loads to off-peak times such as night time and weekends. 
Such demand-side management (DSM) has become important and popular recently because it facilitates the incorporation of renewable energy resources (RES) into the power system by customers.
At the same time, grid operations are significantly impacted by the active participation of customers in electricity dispatch. 
To implement DSM and optimize electricity usage, residential customers often employ home energy management systems (HEMSs). Such HEMSs play a significant role in the energy management of the residential sector and allow the exchange of energy consumption information with the utility to improve the energy profile and reliability of the power grid. 

\begin{figure*}[t]
    \centering
    \includegraphics[width=0.85\linewidth]{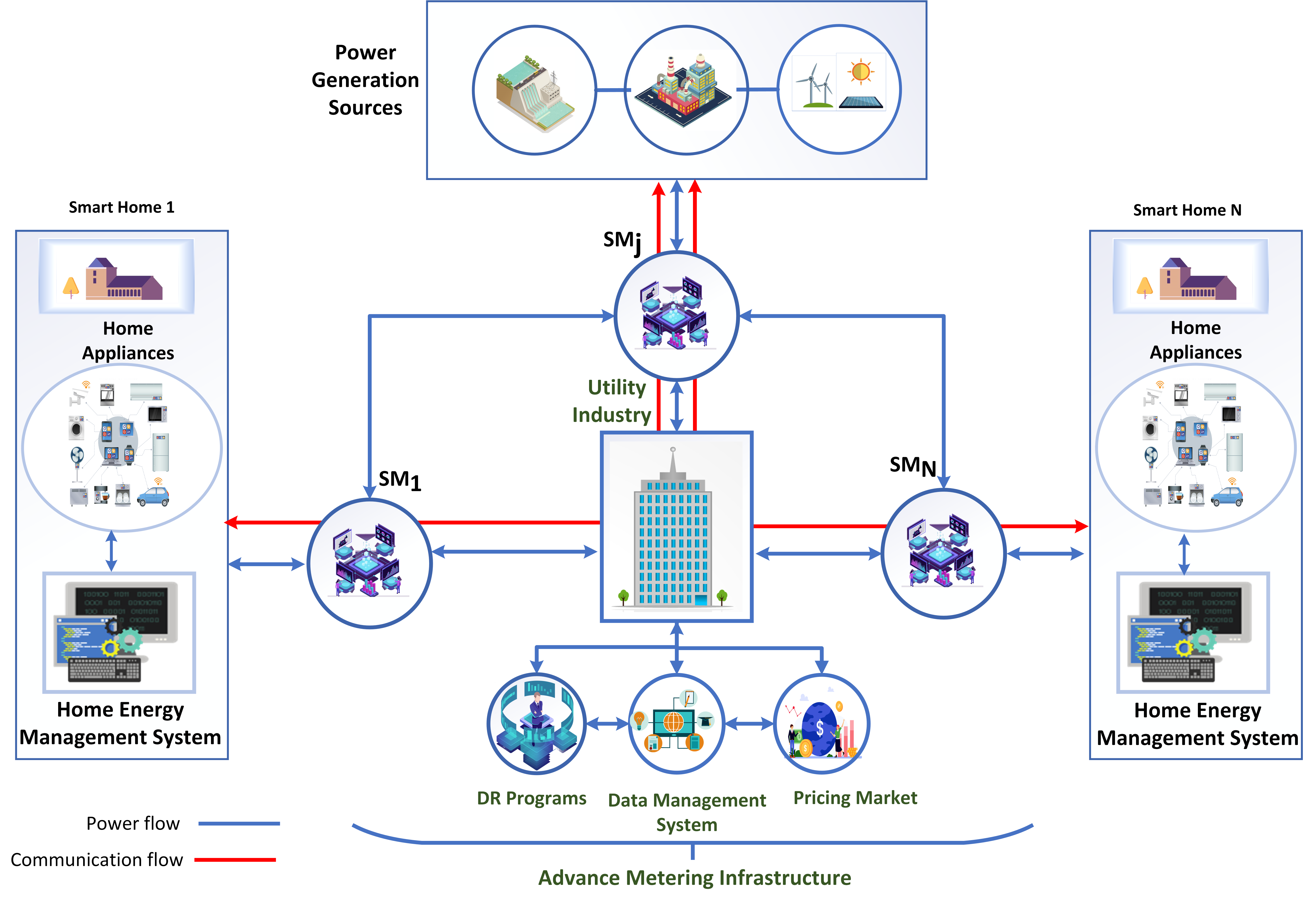}
    \caption{A home energy management system in the electrical power system.}
     \label{fig:HEMS_architecture}
\end{figure*}
An HEMS (Fig. \ref{fig:HEMS_architecture}) is an information and management system to automatically (or semi-automatically) monitor and control the electrical energy production and usage within a household by processing information collected from hardware installed in the electrical power system and the household.
The typical objective of an HEMS is to minimize the customer's costs.
Bidirectional communication among the HEMS, smart meters, the utility, and the power grid enables the HEMS to meet its objectives by, for example, implementing a peak shaving strategy while considering the electricity price signal.
An Internet of Things (IoT) network, along with advanced metering infrastructure (AMI), supports the bidirectional communication and enables robust data management systems, strong network connectivity, and smart metering systems.
The deployment of AMI makes it possible for smart meters to measure and collect useful information, such as energy consumption, available (generated) energy, or the energy price in the next hour, in a precise and timely manner.
Moreover, this information is exchanged between the HEMS and the utility simultaneously in real-time.
As a result, customers can take part in DSM strategies and manage the energy demand effectively.

\begin{figure}[!ht]
    \centering
    \includegraphics[width=\linewidth]{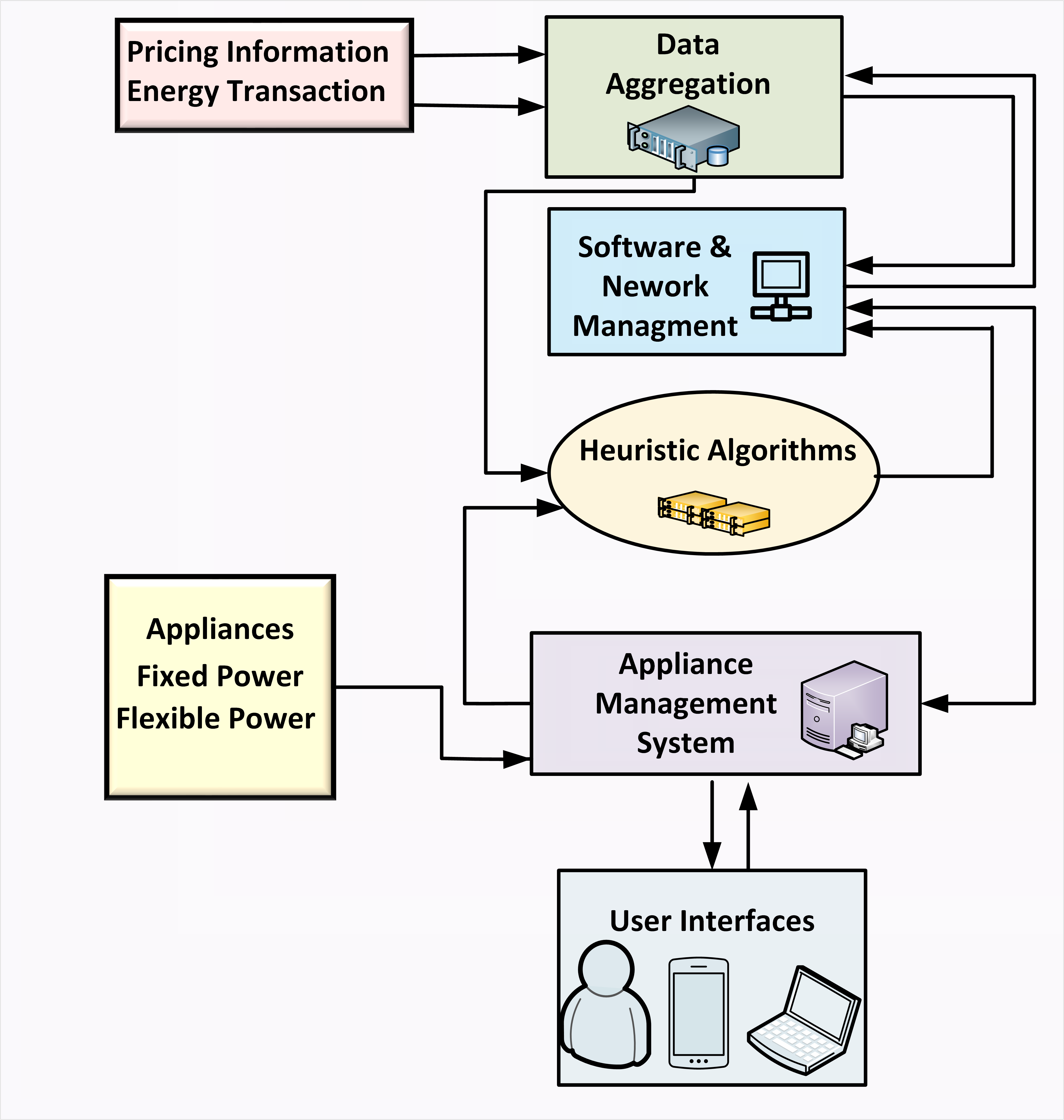}
    \caption{Typical operations of a home energy management system.}
    \label{Fig:HEMS_operation}
\end{figure}

Figure \ref{Fig:HEMS_operation} illustrates the operations of a typical HEMS. Four components---Data Aggregator (DA), Software \& Network Management (SNM), Appliance Management System (AMS), and Heuristic Algorithm (HA) components---interface with each other to form the HEMS. DA receives energy pricing and energy production information and sends them to SNM and HA. The AMS component collects data about appliances, such as energy consumption, operation time interval, and data received from user interfaces etc., and exchanges them with HA and SNM. Thereafter, HA executes the scheduling task and sends the results (new schedule, etc.) to SNM. SNM operates as the primary control and management component, managing the accumulated data of DA, HA, and AMS components and processing the flow of instructions in the network.

%

HEMSs and their characteristics have been extensively investigated in the last decade, and a comprehensive description of HEMS architectures, DSM approaches, smart grid technologies, communication protocols, and various decision-making algorithms can be found in \cite{04}.
This paper focuses on the operational aspects of HEMSs and assesses their resilience against a specific type of cyberattacks.
Such attacks are defined by fake price signals that are used as inputs to the HEMSs to alter their load schedule. 
To the best of the authors' knowledge, this important aspect has not yet been studied in the literature. 
Before we discuss the details of the proposed study presented in Section \ref{cyberattacks}, we will briefly introduce the main ideas behind demand response in Section \ref{subsec:DR} and the scheduling algorithms in Section \ref{sec:he}.

\section{Demand Response}\label{subsec:DR}
The goal of an HEMS is to enable and support DSM to meet specific objectives such as minimization of customers' electricity bills, utility costs, or system costs.
DSM is typically achieved by offering financial incentives to customers, inducing behavioral changes through education, using higher-efficiency loads, increasing diversity factors, using distributed energy resources (DERs), or other measures \cite{Chiu_2013}. 
The continuous integration of RES into the power system has made it important to enable effective DSM in order to match the power supply with the load. 

Demand response (DR) methods, which offer financial incentives to customers, are popular and highly researched techniques to achieve DSM since they incentivize RES integration along with DSM. DR is defined as ``a tariff or program established to motivate changes in electric usage by end-users from their normal consumption patterns in response to changes in the price of electricity over time, or to give incentive payments designed to induce lower electricity use at times of high wholesale market prices or when system reliability is jeopardized'' \cite{02}. 
HEMSs nearly always employ DR methods to achieve their goals. 

DR can be categorized into two types: incentive- and price-based programs \cite{003}.
An incentive-based program involves customers' participation to reallocate their energy consumption in off-peak hours, in response to which a reward (a bill credit or a payment) is given to the customers for their participation.
Incentive-based programs involve direct load control, load curtailing, emergency demand responses, etc. 

On the other hand, a price-based program is a more indirect means of achieving DR. In price-based programs, different pricing signals are sent at different times to the customers. As a result, customers are induced to reduce their energy consumption at certain times in order to take advantage of possible monetary benefits.
Price-based programs include time-of-use (TOU) tariffs, real-time pricing, inclined block rate, critical peak pricing, and day-ahead pricing \cite{03,04,06}. 
In recent researches, price-based DR has been widely studied in the residential sector, particularly in HEMSs.

For price-based DR, the price tariff scheme, i.e., the price bands for different designated time intervals including off-peak, mid-peak, and peak hours, is important.
The TOU tariff scheme is a widely used price tariff scheme in many countries for customers in the residential sector.
The TOU tariff scheme provides the average electricity cost of power generation during different time periods, thereby enabling the customers to manage their energy usage voluntarily.
Customers have flexibility to use electricity either in the peak time interval (which yields a higher cost) or off-peak (lower cost as a result of less stress on the grid).

In this case, DR algorithms depend on the \textit{flexibility} offered by home appliances. An appliance is \textit{flexible} if its energy consumption can be shifted in time within the boundaries of end-user comfort requirements, while maintaining the total consumption \cite{pamela_macdougall_quantifying_2013}.
Home appliances can be divided into two types---\textit{fixed}  and \textit{flexible}---power appliances, based on their characteristics and priorities as follows \cite{24,25}:
\subsubsection{Fixed power appliances}
The appliances in this category have a fixed power consumption profile and operating time, e.g., ceiling fans, lamps,  and TVs.

\subsubsection{Flexible power appliances}
These appliances can be controlled, and their energy consumption profiles can be scheduled by the HEMS.
The operation of flexible power appliances can be controlled by incentive-based 
 or price-based programs.
These loads can be further categorized into two types---uninterruptible and interruptible---depending on whether their operations can be interrupted or not. Table \ref{table_app} lists the appliance classes of fixed and flexible home appliances with their power ratings and operating times \cite{24,25,07}.

\begin{table}[b]
\vspace{1ex}
\centering
\caption{Home appliance characteristics: Type, Power Rating (PR) and Operating time (OT)}
\label{table_app}
\begin{tabular}{l|l|c|c}

\textbf{Appliance}&\textbf{Type} &\textbf{\small PR (kWh)}& \textbf{OT (h)} \\
 \hline
  \hline
         Ceiling fan     & fixed           & 0.075    &14\\ 
         Lamp            & fixed           &  0.1     & 13 \\ 
          TV             & fixed           &0.48      & 7  \\
         Oven            &fixed            &2.3       &6\\
      Washing machine    &flex (Uninterruptible)      & 0.7      & 8\\ 
     Iron                &flex (Uninterruptible)        & 1.8      & 7  \\
       Air conditioner   &flex (Interruptible)        &1.44      & 10 \\ 
      Water heater       &flex (Interruptible)       & 4.45     &8  \\

\hline
\end{tabular}
\end{table}

\section{Heuristic Scheduling Algorithms}\label{HS Algorithms}
\label{sec:he}

Many techniques have been explored to exploit the flexibility in home appliances and to perform DR-based optimization. 
A typical approach is to cleverly adapt optimization techniques to solve linear and non-linear objective functions. 
Recently, artificial intelligence (AI)-based methods have also become popular.
Heuristic scheduling (HS) algorithms comprise an important group of techniques to realize energy optimization and load shifting operations in HEMSs.
Many heuristic algorithms have been explored previously, depending on the problem setup and conditions \cite{04, 05, 06, 07, 08, 10, 11, 12, 13, 14, 15}.
Among the various optimization algorithms, the genetic algorithm (GA) and harmony search algorithm (HSA) are two important algorithms that are particularly suitable for solving constraint-optimization-based scheduling problems and the flexible selection criteria of achieving an optimal (balanced) combination of exploration and exploitation \cite{07,32,33}.
\subsection{Genetic Algorithm} \label{subsec:GA}
GA is a widely applied algorithm due to its fast computational time and easy implementation of many complex problems \cite{holland1992}.
GA is a metaheuristic algorithm inspired by the theory of natural evolution and evolutional processes like genetic inheritance and natural selection,

GA is an iterative process in which a population of potential candidate solutions is first randomly generated.
The population in each iteration is called a generation. 
All the individual candidates (known as genes) in the population are then evaluated using a fitness function (i.e., the problem objective).
The best candidates are stochastically selected from the current generation, and their genome is modified by recombination (crossover) and replacement (mutation) to form a new generation. 
This new generation of candidate solutions is then used in the next iteration.
The stopping criteria for the algorithm are the maximum population size and the  best candidate allocation that satisfies the objective function. 

\subsection{Harmony Search Algorithm} \label{subsec:HSA}

HSA is a popular metaheuristic algorithm inspired by the musical improvisation process \cite{30}. 
Consider a music orchestra that improvises to find and perform the most harmonious and melodious music.
Each musician in an orchestra corresponds to a decision variable, and an instrument's pitch range corresponds to the set of possible values of the decision variable.
The musical harmony produced by the the musicians at a certain time can be considered as the solution vector for an iteration.
An audience's aesthetic judgment of the music can be related to the fitness of the objective function.
Just like a musical orchestra attempts to find (or play) the best music possible by improving it over time, the optimization algorithm aims to progressively find the optimal solution.
Thus, the HSA is an idealized mapping from qualitative improvisation into a quantitative formulation, where musical harmony concepts are applied to an optimization process.  

\subsection{Representative Simulations for Demand Response in Home Energy Management Systems}
%
Some simulation results are now provided to demonstrate the performance of the optimization algorithms GA and HSA.
As the household loads, the eight appliances listed in Table \ref{table_app} are investigated with the given power ratings and operating time periods.
Since the uninterruptible appliances cannot be shifted after they start operating, the HEMS schedules the operation of the iron after the washing machine.
Interruptible appliances, on the other hand, are scheduled based on the pricing signal in any time period. 
The energy consumption of the household appliances for one day (starting from 12 am to 12 am in the following day) with a scheduling resolution of $1$ hour ($t$) is considered.
The TOU pricing tariffs for the summer (May 1 to October 31, 2019) and winter (November 1, 2018 to April 30, 2019) seasons are taken from \cite{28}.

\begin{figure}
  \centering
  \subfloat[Electricity costs in summer season]
  {\includegraphics[width=\linewidth]{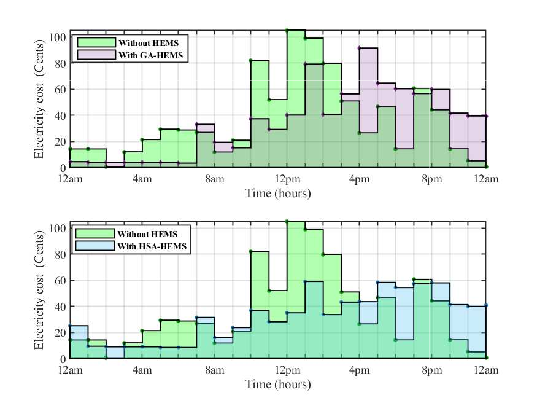}\label{fig:CPH_Summer}}
  \hfill
  \subfloat[Electricity costs in winter season]
  {\includegraphics[width=\linewidth]{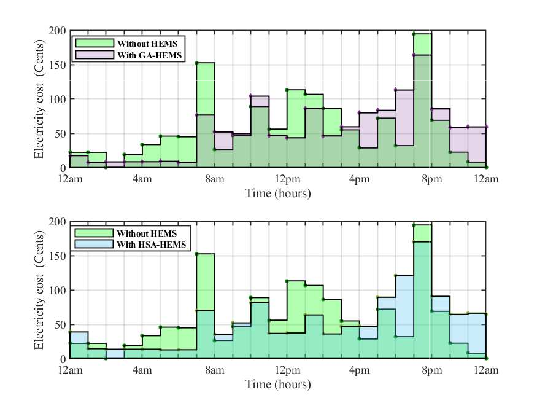}\label{fig:CPH_Winter}}
  \caption{Electricity costs per hour under the time of use pricing scheme for a day each in the summer and winter seasons without and with the two heuristic algorithms, genetic algorithm (GA) and harmony search algorithm (HSA).}
  \label{fig:CPH}
\end{figure}

Figure ~\ref{fig:CPH} presents a comparison of the electricity costs for three cases---``Without HEMS'', and with GA-HEMS and HSA-HEMS---in summer and winter seasons. The deployment of HSA-HEMS and GA-HEMS led to lower total costs as compared to the ``Without HEMS'' case. The energy cost  was highest in the ``Without HEMS'' case, because most of the energy was used either in the peak or mid-peak time. Between the two algorithms, HSA-HEMS reduced the cost by 43.55\% and 11.91\% in the summer and winter seasons, respectively, while GA-HEMS reduced by 23.37\% and 18.91\%, respectively.
%
%

\section{Cyberattacks}\label{cyberattacks}
In a smart grid, the real-time exchange of information, especially data collected from smart meters, electricity pricing markets, and utility companies, requires a secure and protective layer of the communication channel \cite{Hahn_2013}. 
However, the complex structure of the smart grid and the proliferation of smart devices makes it vulnerable to \textit{cyberattacks}. 
A typical cyberattack in a smart grid is the injection of false data into the system to distort the energy demand, grid network states, and electricity pricing signals \cite{Humayed_2017, Aloul_2012}. 
\subsection{How Cyberattacks Work}\label{cyberattacks_working}
Tan et al. \cite{Tan_2013} studied the impact of security threats on a real time pricing system, which could destabilize the electricity market or even cause severe failures.
They delineated defensive measures against two classes of data integrity attacks: scaling (the meter reads an amplified version of the actual prices) and delaying (the meter uses old prices).
In \cite{Giraldo_2016}, the authors systemically examined the arbitrary injection of pricing (data) signals and proposed countermeasures based on a cumulative sum control chart (CUSUM) technique to identify the attacks. 
The injection of false data creates a disparity between the generated and consumed power, which subsequently leads to two major problems: (i) instability of the entire system, and (ii) increase in the operational costs by the addition of forged data to the electricity market\cite{Rawat_2015,Xie_2011,liu2015}.
\begin{figure}
    \centering
    \includegraphics[width=\linewidth]{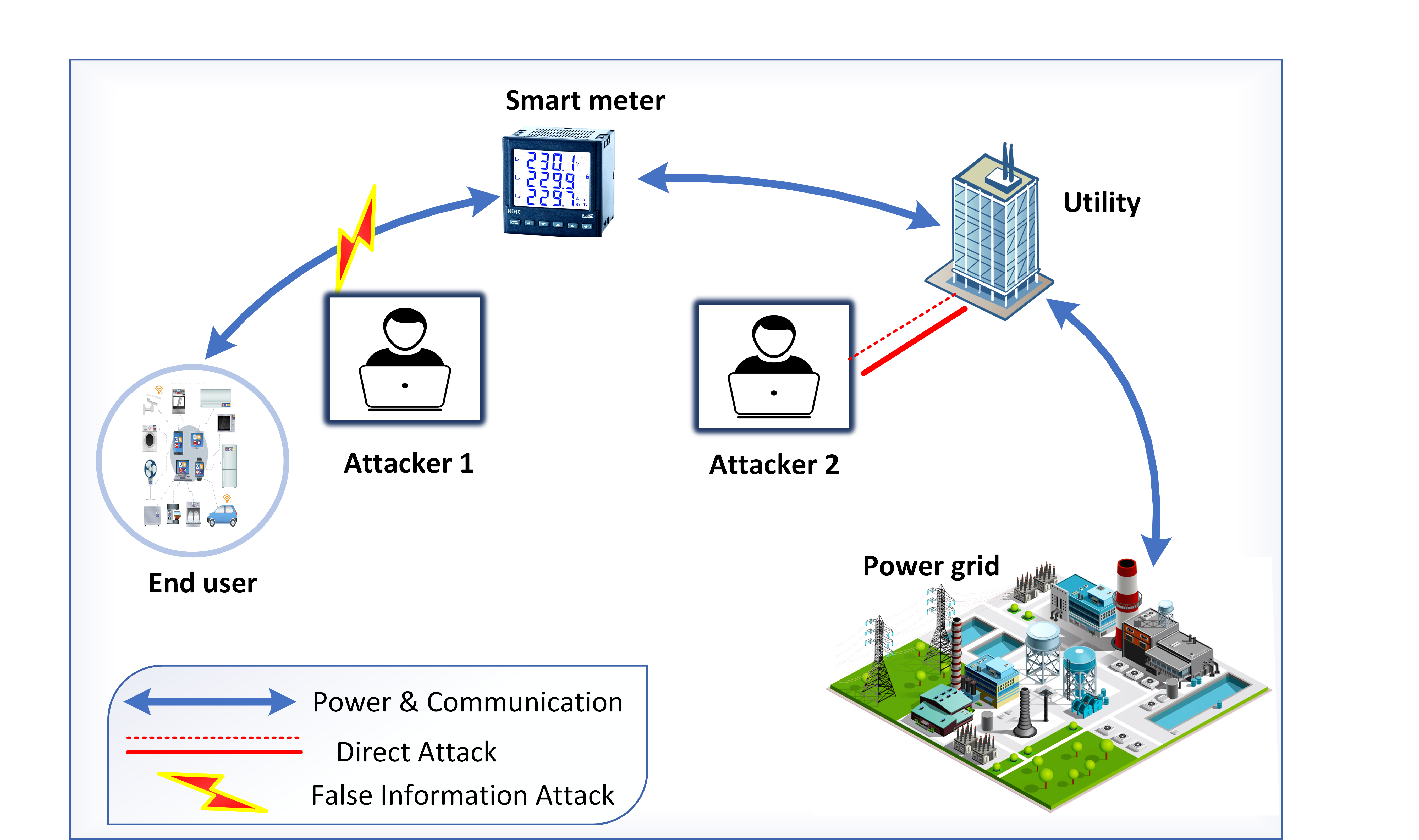}
    \caption{Cyberattack on a cyber physical system: Attacker 1  attempts to inject wrong pricing data or alter energy demand information, and Attacker 2 attempts a direct attack on the utility to alter the energy demand and/or production.}
     \label{fig:cyberattack}
\end{figure}

Figure \ref{fig:cyberattack} depicts possible cyberattacks on a cyber physical system comprising the communication infrastructure of various components associated with a smart grid connected to an end-user household.
The utility collects information related to the energy demand (generation, consumption, and price) through the AMI and transmits this information to the smart meters and end users through an IoT or WiFi network.

The hierarchical communication infrastructure shown above is exposed to three kinds of cyberattacks \cite{liu2015}. 
First, an adversary can attack the utility main system (computer devices) and change the pricing curve. 
Subsequently, this information is sent to the end users, and based on the fake price, the HEMS schedules their loads. 
Secondly, an attacker can directly attack smart meters at or near the end-user household and tamper with the received (or transmitted) data.
An adversary can also attack any access point in the WiFi network, create a (fake) access point, and send false pricing data to the smart meter. 
%


\subsection{Cyberattack Scenario}
Consider an HEMS that employs HS  algorithms to perform DR-based optimization based on the received price signals. 
As discussed in \cite{Giraldo_2016}, a smart meter or other receivers can often be hacked with minimum effort due to the lack of security measures.
Let us examine a scenario where an attacker has the resources to hack into a smart meter and inject corrupted (price) information.
The cyberattacker aims to mislead the heuristics to induce a higher electricity bill or peak demand by modifying the peak prices arbitrarily, which increases the mismatch between the generated energy and the energy demand. 
For example, in the case of TOU tariffs in winter, the peak time prices of 20.8 cent/kWh occur from 7 am to 11 am and from 6 pm to 8 am \cite{30}. 
The attacker can now alter these peak prices either by shifting them to the off-peak time or by simply directly lowering the prices, which, in turn, increases/decreases the electricity bill.

\begin{figure}
  \centering
  \subfloat[Electricity costs per hour under the cyberattack case]
  {\includegraphics[width=\linewidth]{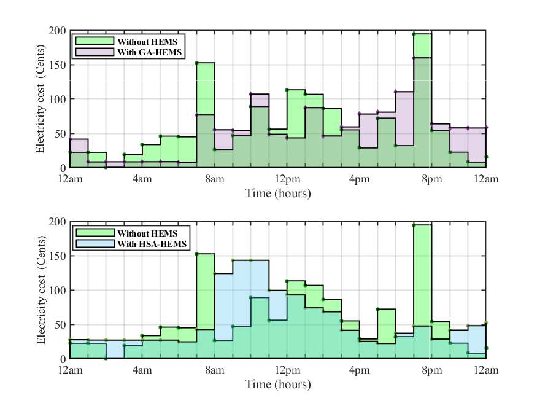}\label{fig:CPH_attack}}
  \hfill
  \subfloat[Resilience of GA-HEMS and HSA-HEMS---represented by a resilience index]
  {\includegraphics[width=\linewidth]{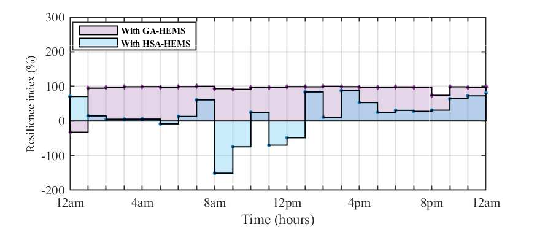}\label{fig:RI}}
  \caption{The impact on the electricity costs for a day when a cyberattacker changes the time of use pricing. Two heuristic optimization algorithms, genetic algorithm (GA) and harmony search algorithm (HSA), optimize the costs for the home energy management system (HEMS).}
\end{figure}

In such a scenario, how do the designed models using GA and HSA algorithms react when the system is attacked and forged pricing information is injected? To analyze this, assume that the adversary particularly targets the peak prices of the energy demand, i.e., from 7 am to 11 am and from 6 am to 8 am. 
Figure \ref{fig:CPH_attack} presents the electricity costs for a day in winter after a cyberattack has occurred. 
The GA-HEMS and HSA-HEMS attempt to schedule the energy consumption as before, but the electricity costs naturally increase.
However, this increase is not very high.
The GA increases the cost by 0.15\% as compared to the optimal cost achieved earlier without the cyberattack, whereas the HSA increases the cost by 1.8\%.

The resilience of any algorithm against cyberattacks can be characterized by measuring how much the forged pricing data affects the performance of the considered system metrics (here, electricity costs) in the designed scenario. A simple way to measure the resilience is by using a resilience index (RI) as follows:
{\begin{equation}
\text{RI} = 100 - \Bigl(\frac{|{C_A} - {C}_O|}{{C_O}}\Bigr)\times100
\end{equation}}
Here, ${C_A}$  and ${C_O}$ represent the total electricity cost when the system is under attack and otherwise, respectively. In both cases, the total cost is optimized using the HEMS. 
Thus, the RI gives a measure of accuracy of the heuristic algorithm against cyberattacks. 
$\text{RI}\in[-\infty\;100\%]$. $\text{RI} = 100\%$ means that the algorithm is extremely resilient ($C_A = C_O$).
As the amount of deviation from the optimal cost increases, RI decreases from the maximum of 100\%. RI becomes negative when $C_A > 2C_O$. 
Negative RI means that the algorithm's performance is poor; the new cost is more than twice the actual cost. 
%

Figure \ref{fig:RI} presents the RI for the designed model for a day. GA-HEMS maintains a good and somewhat constant RI across the day, whereas the HSA has poor RI some times.
Further, the overall RI for GA and HSA for the entire day was 99.8\% and 97.8\%, respectively. 
Thus, even though the cyberattacker attempts to mislead the designed heuristic approaches with fake price information, both the designed algorithms perform robustly against these attacks, providing a similar performance to the case without active management.

\section{Conclusions: Cyberattacks and Future Power Systems}
Modern power systems (MPSs) have added flexibility and coordination by utilizing information and communication technologies (ICT) and AMI.
MPSs have now gradually transitioned into a complex cyber-physical energy system (CPES). 
The cyber layer has not only made it possible for MPSs to become more responsive to faults and other systemic problems but also to co-ordinate production and load energy by reacting faster and smarter to changes. 
Moreover, individual households are empowered to install HEMS to manage their own production and load as well as interactions with the power system.
The efficient transformation of an MPS into a CPES is doubly important today because global climate-change issues have made it necessary to integrate large amounts of RES into the power system.

However, this transformation comes with a price: \textit{vulnerability to cyberattacks}. 
MPS control and operations are more visible to external actors, and the strong interactions between the cyber-physical layers in CPES increase the MPS's vulnerability to cyberattacks.
Moreover, power electronic converters, which are key enablers for integrating RES into MPSs, are typically controlled by employing a hierarchical three-stage structure, namely, primary, secondary and tertiary layers.
This means that the MPSs have additional vulnerabilities and possible attack points in different layers of the system. 
A cyberattacker can take advantage of any software flaws or failures in any layer of the CPES and create harmful disturbances in the system. 

How an MPS will deal with such cyberattacks in the future will be critical to ensure its stability and performance. 
Advanced and resilient technologies and mitigation measures has to be developed and implemented at every level.  
Hierarchical stages in MPSs enforce different timescales of operation, giving great flexibility to design mitigation techniques against cyberattacks.
At the same time, these measures can also be cheated, if the attacker has access to multiple points to design \textit{coordinated} attacks \cite{subham}.
Data-driven techniques are a computationally viable platform to identify such anomalies. 
Robust and resilient control strategies using watermarking \cite{wee} and state observers \cite{pas} could be smartly employed to infiltrate such cyber attacks in the primary and secondary control layer by guaranteeing faster action.

For researchers and industry practitioners, the development of countermeasures to mitigate the impacts of cyberattacks, including financial losses, privacy invasions, data losses, etc., is a fascinating and highly relevant area of investigation today. After all, a safe and secure electrical power system is an important part of a safe and secure society.

\bibliographystyle{IEEEtran}
\bibliography{ref}

\begin{thebibliography}{10}
\providecommand{\url}[1]{#1}
\csname url@samestyle\endcsname
\providecommand{\newblock}{\relax}
\providecommand{\bibinfo}[2]{#2}
\providecommand{\BIBentrySTDinterwordspacing}{\spaceskip=0pt\relax}
\providecommand{\BIBentryALTinterwordstretchfactor}{4}
\providecommand{\BIBentryALTinterwordspacing}{\spaceskip=\fontdimen2\font plus
\BIBentryALTinterwordstretchfactor\fontdimen3\font minus
  \fontdimen4\font\relax}
\providecommand{\BIBforeignlanguage}[2]{{%
\expandafter\ifx\csname l@#1\endcsname\relax
\typeout{** WARNING: IEEEtran.bst: No hyphenation pattern has been}%
\typeout{** loaded for the language `#1'. Using the pattern for}%
\typeout{** the default language instead.}%
\else
\language=\csname l@#1\endcsname
\fi
#2}}
\providecommand{\BIBdecl}{\relax}
\BIBdecl

\bibitem{01}
R.~Lu, S.~H. Hong, and M.~Yu, ``Demand response for home energy management
  using reinforcement learning and artificial neural network,'' \emph{IEEE
  Transactions on Smart Grid}, 2019.

\bibitem{04}
H.~Shareef, M.~S. Ahmed, A.~Mohamed, and E.~Al~Hassan, ``Review on home energy
  management system considering demand responses, smart technologies, and
  intelligent controllers,'' \emph{IEEE Access}, vol.~6, pp. 24\,498--24\,509,
  2018.

\bibitem{Chiu_2013}
W.~{Chiu}, H.~{Sun}, and H.~V. {Poor}, ``Energy imbalance management using a
  robust pricing scheme,'' \emph{IEEE Transactions on Smart Grid}, vol.~4,
  no.~2, pp. 896--904, 2013.

\bibitem{02}
Q.~Qdr, ``Benefits of demand response in electricity markets and
  recommendations for achieving them,'' \emph{US Dept. Energy, Washington, DC,
  USA, Tech. Rep}, 2006.

\bibitem{003}
P.~Siano, ``Demand response and smart grids—a survey,'' \emph{Renewable and
  sustainable energy reviews}, vol.~30, pp. 461--478, 2014.

\bibitem{03}
J.~S. Vardakas, N.~Zorba, and C.~V. Verikoukis, ``A survey on demand response
  programs in smart grids: Pricing methods and optimization algorithms,''
  \emph{IEEE Communications Surveys \& Tutorials}, vol.~17, no.~1, pp.
  152--178, 2014.

\bibitem{06}
W.~Fan, N.~Liu, and J.~Zhang, ``Multi-objective optimization model for energy
  mangement of household micro-grids participating in demand response,'' in
  \emph{2015 IEEE Innovative Smart Grid Technologies-Asia (ISGT ASIA)}.\hskip
  1em plus 0.5em minus 0.4em\relax IEEE, 2015, pp. 1--6.

\bibitem{pamela_macdougall_quantifying_2013}
{Pamela MacDougall}, {Bart Roossien}, {Cor Warmer}, and {Koen Kok},
  ``Quantifying flexibility for smart grid services,'' in \emph{Power and
  {{Energy Society General Meeting}} ({{PES}}), 2013 {{IEEE}}}, {Vancouver,
  Canada}, Jul. 2013, pp. 1--5.

\bibitem{24}
H.~Hussain, N.~Javaid, S.~Iqbal, Q.~Hasan, K.~Aurangzeb, and M.~Alhussein, ``An
  efficient demand side management system with a new optimized home energy
  management controller in smart grid,'' \emph{Energies}, vol.~11, no.~1, p.
  190, 2018.

\bibitem{25}
Y.~Liu, C.~Yuen, R.~Yu, Y.~Zhang, and S.~Xie, ``Queuing-based energy
  consumption management for heterogeneous residential demands in smart grid,''
  \emph{IEEE Transactions on Smart Grid}, vol.~7, no.~3, pp. 1650--1659, 2015.

\bibitem{07}
Q.~Zhu, J.~Zhang, Y.~Hou, and Y.~Qiao, ``The energy-saving scheduling of campus
  classrooms: A simulation model,'' \emph{IEEE Systems, Man, and Cybernetics
  Magazine}, vol.~7, no.~2, pp. 22--34, 2021.

\bibitem{05}
B.~Gunay and W.~Shen, ``Connected and distributed sensing in buildings:
  Improving operation and maintenance,'' \emph{IEEE Systems, Man, and
  Cybernetics Magazine}, vol.~3, no.~4, pp. 27--34, 2017.

\bibitem{08}
H.~M. Hussain and P.~H. Nardelli, ``A heuristic-based home energy management
  system for demand response,'' in \emph{2020 IEEE Conference on Industrial
  Cyberphysical Systems (ICPS)}, vol.~1.\hskip 1em plus 0.5em minus 0.4em\relax
  IEEE, 2020, pp. 285--290.

\bibitem{10}
P.~Nardelli, H.~M. Hussain, A.~Narayanan, and Y.~Yang, ``Virtual microgrid
  management via software-defined energy network for electricity sharing:
  Benefits and challenges,'' \emph{IEEE Systems, Man, and Cybernetics
  Magazine}, vol.~7, no.~3, pp. 10--19, 2021.

\bibitem{11}
Z.~Zhao \emph{et~al.}, ``An optimal power scheduling method for demand response
  in home energy management system,'' \emph{IEEE Transactions on Smart Grid},
  vol.~4, no.~3, pp. 1391--1400, 2013.

\bibitem{12}
C.~Bharathi, D.~Rekha, and V.~Vijayakumar, ``Genetic algorithm based demand
  side management for smart grid,'' \emph{Wireless Personal Communications},
  vol.~93, no.~2, pp. 481--502, 2017.

\bibitem{13}
M.~S. Ahmed, A.~Mohamed, T.~Khatib, H.~Shareef, R.~Z. Homod, and J.~A. Ali,
  ``Real time optimal schedule controller for home energy management system
  using new binary backtracking search algorithm,'' \emph{Energy and
  Buildings}, vol. 138, pp. 215--227, 2017.

\bibitem{14}
K.~Ma, T.~Yao, J.~Yang, and X.~Guan, ``Residential power scheduling for demand
  response in smart grid,'' \emph{International Journal of Electrical Power \&
  Energy Systems}, vol.~78, pp. 320--325, 2016.

\bibitem{15}
P.~H. Nardelli and F.~K{\"u}hnlenz, ``Why smart appliances may result in a
  stupid grid: Examining the layers of the sociotechnical systems,'' \emph{IEEE
  Systems, Man, and Cybernetics Magazine}, vol.~4, no.~4, pp. 21--27, 2018.

\bibitem{32}
X.~Z. Gao, V.~Govindasamy, H.~Xu, X.~Wang, and K.~Zenger, ``Harmony search
  method: theory and applications,'' \emph{Computational intelligence and
  neuroscience}, vol. 2015, 2015.

\bibitem{33}
T.~El-Ghazali, ``Metaheuristics: from design to implementation,'' \emph{Jonh
  Wiley and Sons Inc., Chichester}, vol.~9, pp. 10--11, 2009.

\bibitem{holland1992}
J.~H. Holland \emph{et~al.}, \emph{Adaptation in natural and artificial
  systems: an introductory analysis with applications to biology, control, and
  artificial intelligence}.\hskip 1em plus 0.5em minus 0.4em\relax MIT press,
  1992.

\bibitem{30}
O.~Abdel-Raouf and M.~A.-B. Metwally, ``A survey of harmony search algorithm,''
  \emph{International Journal of Computer Applications}, vol.~70, no.~28, 2013.

\bibitem{28}
``{Time-of-Use (TOU) Pricing and Schedules},''
  \url{https://www.powerstream.ca/customers/rates-support-programs/time-of-use-pricing.html},
  2019.

\bibitem{Hahn_2013}
A.~Hahn, A.~Ashok, S.~Sridhar, and M.~Govindarasu, ``Cyber-physical security
  testbeds: Architecture, application, and evaluation for smart grid,''
  \emph{IEEE Transactions on Smart Grid}, vol.~4, no.~2, pp. 847--855, 2013.

\bibitem{Humayed_2017}
A.~Humayed, J.~Lin, F.~Li, and B.~Luo, ``Cyber-physical systems security—a
  survey,'' \emph{IEEE Internet of Things Journal}, vol.~4, no.~6, pp.
  1802--1831, 2017.

\bibitem{Aloul_2012}
F.~Aloul, A.~Al-Ali, R.~Al-Dalky, M.~Al-Mardini, and W.~El-Hajj, ``Smart grid
  security: Threats, vulnerabilities and solutions,'' \emph{International
  Journal of Smart Grid and Clean Energy}, vol.~1, no.~1, pp. 1--6, 2012.

\bibitem{Tan_2013}
R.~Tan, V.~Badrinath~Krishna, D.~K. Yau, and Z.~Kalbarczyk, ``Impact of
  integrity attacks on real-time pricing in smart grids,'' in \emph{Proceedings
  of the 2013 ACM SIGSAC conference on Computer \& communications
  security}.\hskip 1em plus 0.5em minus 0.4em\relax ACM, 2013, pp. 439--450.

\bibitem{Giraldo_2016}
J.~Giraldo, A.~C{\'a}rdenas, and N.~Quijano, ``Integrity attacks on real-time
  pricing in smart grids: impact and countermeasures,'' \emph{IEEE Transactions
  on Smart Grid}, vol.~8, no.~5, pp. 2249--2257, 2016.

\bibitem{Rawat_2015}
D.~B. Rawat and C.~Bajracharya, ``Detection of false data injection attacks in
  smart grid communication systems,'' \emph{IEEE Signal Processing Letters},
  vol.~22, no.~10, pp. 1652--1656, 2015.

\bibitem{Xie_2011}
L.~Xie, Y.~Mo, and B.~Sinopoli, ``Integrity data attacks in power market
  operations,'' \emph{IEEE Transactions on Smart Grid}, vol.~2, no.~4, pp.
  659--666, 2011.

\bibitem{liu2015}
Y.~Liu, S.~Hu, and T.-Y. Ho, ``Leveraging strategic detection techniques for
  smart home pricing cyberattacks,'' \emph{IEEE Transactions on Dependable and
  Secure Computing}, vol.~13, no.~2, pp. 220--235, 2015.

\bibitem{subham}
S.~Sahoo, T.~Dragi{\v{c}}evi{\'c}, and F.~Blaabjerg, ``Cyber security in
  control of grid-tied power electronic converters--challenges and
  vulnerabilities,'' \emph{IEEE Journal of Emerging and Selected Topics in
  Power Electronics}, 2019.

\bibitem{wee}
S.~Weerakkody, Y.~Mo, and B.~Sinopoli, ``Detecting integrity attacks on control
  systems using robust physical watermarking,'' in \emph{53rd IEEE Conference
  on Decision and Control}.\hskip 1em plus 0.5em minus 0.4em\relax IEEE, 2014,
  pp. 3757--3764.

\bibitem{pas}
F.~Pasqualetti, F.~Dorfler, and F.~Bullo, ``Control-theoretic methods for
  cyberphysical security: Geometric principles for optimal cross-layer
  resilient control systems,'' \emph{IEEE Control Systems Magazine}, vol.~35,
  no.~1, pp. 110--127, 2015.

\end{thebibliography}

\begin{IEEEbiography}[{\includegraphics[width=1in,height=1.25in,clip,keepaspectratio]{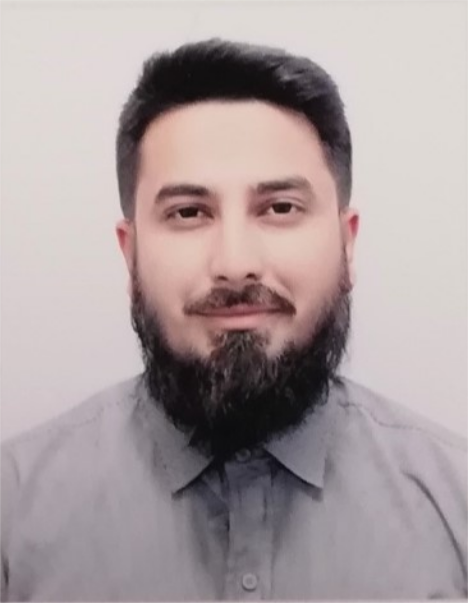}}]{Hafiz Majid Hussain} (M'19) completed his BS and MS in Electrical Engineering from the National University of Computer \& Emerging Sciences and University of Engineering \& Technology Taxila, Pakistan, in 2014 and 2017, respectively. He is the part of the project called Building the Energy Internet as a large-scale IoT-based cyberphysical system. Currently, he is pursuing a Ph.D. towards Electrical Engineering from the Lappeenranta University of Technology in the research group Cyber-Physical Systems Group, Finland. His research interest includes demand response applications, energy resource optimization in smart grid, and information security technologies. More information: https://sites.google.com/view/hafizmajidhussain/biography
\end{IEEEbiography}
\begin{IEEEbiography}[{\includegraphics[width=1in,height=1.25in,clip,keepaspectratio]{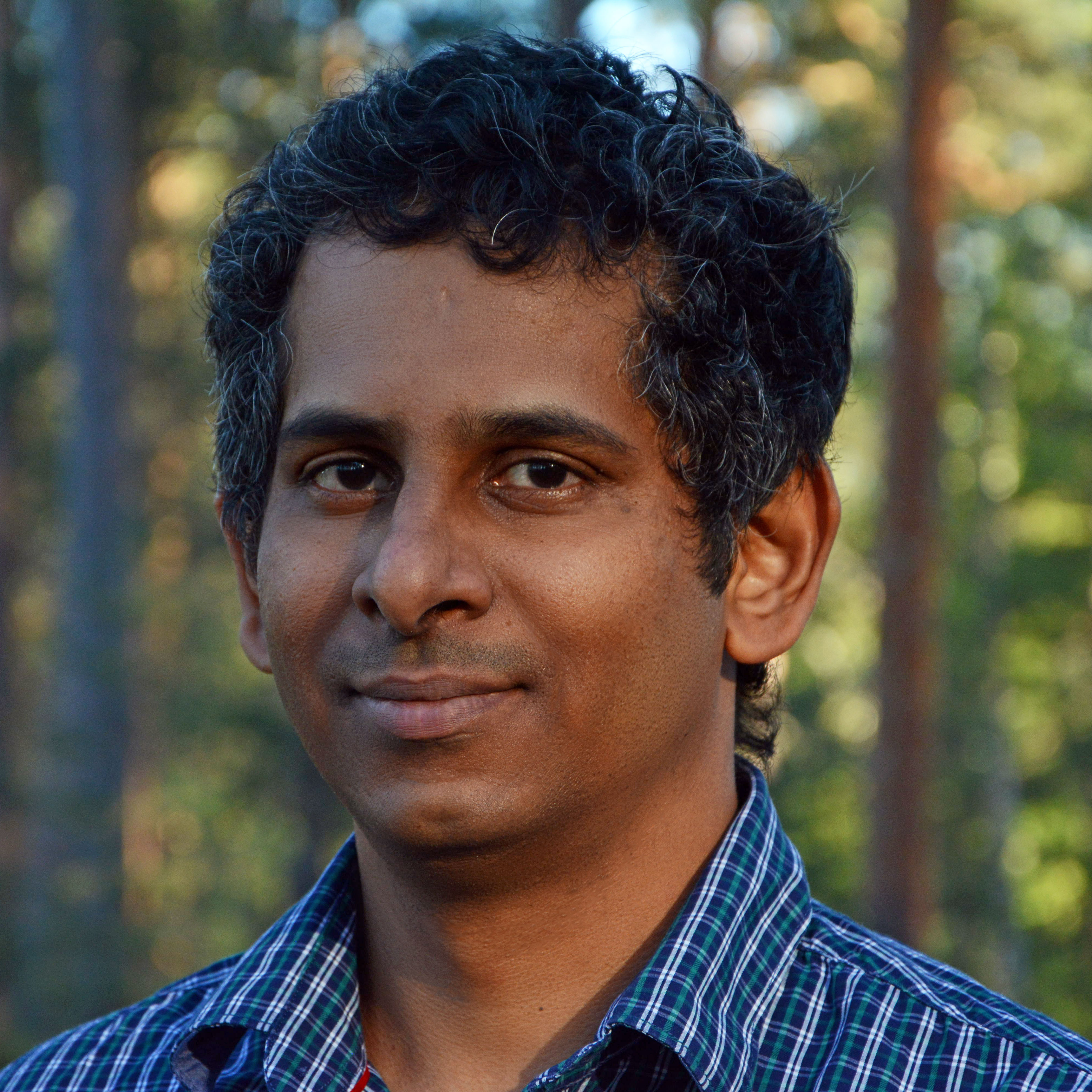}}]{Arun Narayanan} (M'14) received his B.E. degree in Electrical Engineering from Visvesvaraya National Institute of Technology, Nagpur, India and M.Sc. in Energy Technology from Lappeenranta University of Technology (LUT), Finland, in 2002 and 2013, respectively. He subsequently completed his Ph.D. from the School of Energy Systems, LUT University. 
He is currently a Postdoctoral Fellow with LUT University, Lappeenranta, Finland, in the research group Cyber-Physical Systems Group. His research interests include renewable energy-based smart microgrids, electricity markets, demand-side management, energy management systems, and information and communications technology. He focuses on applying optimization, computational concepts, and artificial intelligence techniques to renewable electrical energy problems.
\end{IEEEbiography}
\begin{IEEEbiography}[{\includegraphics[width=1in,height=1.25in,clip,keepaspectratio]{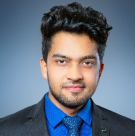}}]
{Subham Sahoo} (S’16-M’18) received the B.Tech. \& Ph.D. degree in Electrical and Electronics Engineering from VSS University of Technology, Burla, India and Electrical Engineering at Indian Institute of Technology, Delhi, New Delhi, India in 2014 \& 2018, respectively. He has worked as a Visiting Student with the Department of Electrical and Electronics Engineering in Cardiff University, UK in 2017. Prior to completion of his PhD, he worked as a postdoctoral researcher in the Department of Electrical and Computer Engineering in National University of Singapore during 2018-19 and in Aalborg University (AAU), Denmark during 2019-2020. He is currently an Assistant Professor in the Department of Energy, AAU, Denmark.

He is a recipient of the Indian National Academy of Engineering (INAE) Innovative Students Project Award for his PhD thesis across all the institutes in India for the year 2019. He was also a distinguished reviewer for IEEE Transactions on Smart Grid in the year 2020. He currently serves as a secretary of IEEE Young Professionals Affinity Group, Denmark and Joint IAS/IES/PELS in Denmark section. 
His research interests are control, optimization, and stability of power electronic dominated grids, renewable energy integration, physics-informed AI tools for cyber-physical power electronic systems.
\end{IEEEbiography}
\begin{IEEEbiography}[{\includegraphics[width=1in,height=1.25in,clip,keepaspectratio]{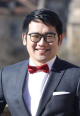}}]{Yongheng Yang}(SM’17) received the B.Eng. degree in Electrical Engineering and Automation from Northwestern Polytechnical University, China, in 2009 and the Ph.D. degree in Energy Technology (power electronics and drives) from Aalborg University, Denmark, in 2014. 
He was a postgraduate student with Southeast University, China, from 2009 to 2011. In 2013, he spent three months as a Visiting Scholar at Texas A\&M University, USA. Since 2014, he has been with the Department of Energy, Aalborg University, where he became a tenured Associate Professor in 2018. In January 2021, he joined Zhejiang University, China, where he is currently a ZJU100 Professor with the Institute of Power Electronics, College of Electrical Engineering. His current research interests include the grid-integration of photovoltaic systems and control of power converters, in particular, the mechanism and control of grid-forming power converters and systems.

Dr. Yang was the Chair of the IEEE Denmark Section (2019-2020). He is an Associate Editor for several IEEE Transactions/Journals. He is a Deputy Editor of the IET Renewable Power Generation for Solar Photovoltaic Systems. He was the recipient of the 2018 IET Renewable Power Generation Premium Award and was an Outstanding Reviewer for the IEEE TRANSACTIONS ON POWER ELECTRONICS in 2018. He received the 2021 Richard M. Bass Outstanding Young Power Electronics Engineer Award from the IEEE Power Electronics Society (PELS). In addition, he has received two IEEE Best Paper Awards. He is currently the Secretary of the IEEE PELS Technical Committee on Sustainable Energy Systems.
\end{IEEEbiography}
\begin{IEEEbiography}[{\includegraphics[width=1in,height=1.25in,clip,keepaspectratio]{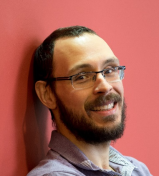}}]{Pedro H. J. Nardelli} received the B.S. and M.Sc. degrees in electrical engineering from the State University of Campinas, Brazil, in 2006 and 2008, respectively. In 2013, he received his doctoral degree from University of Oulu, Finland, and State University of Campinas following a dual degree agreement. He is currently Associate Professor (tenure track) in IoT in Energy Systems at LUT University, Finland, and holds a position of Academy of Finland Research Fellow with a project called Building the Energy Internet as a large-scale IoT-based cyber-physical system that manages the energy inventory of distribution grids as discretized packets via machine-type communications (EnergyNet). He leads the Cyber-Physical Systems Group at LUT and is Project Coordinator of the CHIST-ERA European consortium Framework for the Identification of Rare Events via Machine Learning and IoT Networks (FIREMAN). He is also Docent at University of Oulu in the topic of “communications strategies and information processing in energy systems”. His research focuses on wireless communications particularly applied in industrial automation and energy systems. He received a best paper award of IEEE PES Innovative Smart Grid Technologies Latin America 2019 in the track “Big Data and Internet of Things”. He is also IEEE Senior Member. More information: https://sites.google.com/view/nardelli/
\end{IEEEbiography}
\begin{IEEEbiography}[{\includegraphics[width=1in,height=1.25in,clip,keepaspectratio]{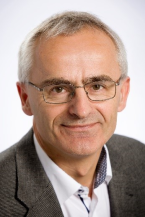}}]
{Frede Blaabjerg} (S’86–M’88–SM’97–F’03) was with ABB-Scandia, Randers, Denmark, from 1987 to 1988. From 1988 to 1992, he got a Ph.D. degree in Electrical Engineering at Aalborg University in 1995. He became an Assistant Professor in 1992, an Associate Professor in 1996, and a Full Professor of power electronics and drives in 1998. From 2017 he became a Villum Investigator. He is honoris causa at University Politehnica Timisoara (UPT), Romania, and Tallinn Technical University (TTU) in Estonia. His current research interests include power electronics and its applications, such as in wind turbines, PV systems, reliability, harmonics, and adjustable speed drives. He has published more than 600 journal papers in the fields of power electronics and its applications. He is the co-author of four monographs and editor of ten books in power electronics and its applications. 

He has received 32 IEEE Prize Paper Awards, the IEEE PELS Distinguished Service Award in 2009, the EPE-PEMC Council Award in 2010, the IEEE William E. Newell Power Electronics Award 2014, the Villum Kann Rasmussen Research Award 2014, the Global Energy uPrize in 2019 and the 2020 IEEE Edison Medal. He was the Editor-in-Chief of the IEEE TRANSACTIONS ON POWER ELECTRONICS from 2006 to 2012. He has been a Distinguished Lecturer for the IEEE Power Electronics Society from 2005 to 2007 and for the IEEE Industry Applications Society from 2010 to 2011 as well as 2017 to 2018. In 2019-2020 he served a President of the IEEE Power Electronics Society. He is Vice-President of the Danish Academy of Technical Sciences too. He is nominated in 2014-2019 by Thomson Reuters to be between the most 250 cited researchers in Engineering in the world.
\end{IEEEbiography}

\end{document}